# SARS-CoV-2 Virus-Like Particles with Plasmonic Au Cores and S1-Spike Protein Coronas

Weronika Andrzejewska,[1] Barbara Peplińska,[1] Jagoda Litowczenko,[1] Patryk Obstarczyk,[2] Joanna Olesiak-Bańska,[2] Stefan Jurga,[1] Mikołaj Lewandowski[1,*]

*[1]NanoBioMedical Centre, Adam Mickiewicz University, Wszechnicy Piastowskiej 3, 61-614 Poznań, Poland*

*[2]Institute of Advanced Materials, Wroclaw University of Science and Technology, Wybrzeże Wyspiańskiego 2, 50-370 Wrocław, Poland*

*[*]Corresponding author: lewandowski@amu.edu.pl*

**Abstract**

The COVID-19 pandemic has stimulated the scientific world to intensify virus-related studies, aimed at the development of quick and safe ways of detecting viruses in human body, studying the virus-antibody and virus-cell interactions, and designing nanocarriers for targeted antiviral therapies. However, research on dangerous viruses can only be performed in certified laboratories that follow strict safety procedures. Thus, developing deactivated virus constructs or safe-to-use virus-like objects, which imitate real viruses and allow performing virus-related studies in any research laboratory, constitutes an important scientific challenge. One of the groups of such species are the so-called virus-like particles (VLPs). Instead of capsids with viral DNA/RNA, VLPs have synthetic cores with real virus proteins attached to them. We have developed a method for the preparation of VLPs imitating the virus responsible for the COVID-19 disease: the SARS-CoV-2. The particles have Au cores surrounded by "coronas" of S1 domains of the virus's Spike protein. Importantly, they are safe to use and specifically interact with SARS-CoV-2 antibodies. Moreover, Au cores exhibit localized surface plasmon resonance (LSPR), which makes the synthesized VLPs suitable for biosensing applications. Within our studies, the effect allowed us to visualize the interaction between the VLPs and the antibodies and identify the characteristic vibrational signals. What is more, additional functionalization of the particles with a fluorescent label revealed their potential in studying specific virus-related interactions. Notably, the universal character of the developed synthesis method makes it potentially applicable for fabricating VLPs imitating other life-threatening viruses.

## 1. Introduction

Since the beginning of the COVID-19 (Coronavirus Disease 2019) pandemic, scientists around the world have been intensively studying the virus responsible for the disease: the SARS-CoV-2 (Severe Acute Respiratory Syndrome Coronavirus 2), trying to determine its structure and biological properties (such as the molecular mechanisms of human infection, cellular targets and life cycle).[1] Gaining this information is necessary for the development of effective and rapid virus detection methods at the early stage of infection, as well as inventing new medicines and new generation vaccines. However, conducting research with the use of infectious viral particles (even with inactivated capsids) is related to a potential health risk. Therefore, such studies can only be performed in scientific laboratories with the highest class biosafety (biological safety levels 3 and 4).[2,3] This problem is addressed by the idea of using virus-like particles (VLPs) – non-infectious and safe to use biomimetic species that resemble certain features of a real viral molecule.[4] VLPs can be used in vaccines, serve as virus phantoms, vehicles for targeted delivery of different biological materials (genes, peptides, drugs) and bioimaging contrast agents.[5] One type of VLPs are those consisting of metallic cores and the surrounding "protein coronas".[6] In this case, the cores usually are characterized by potentially-applicable physical and chemical properties, while the coronas constitute bioactive layers and reduce the surface free energy of the cores.

In this work, we describe the methodology of synthesizing VLPs imitating the SARS-CoV-2. Au nanoparticles (AuNPs) of a size of ~90-100 nm – similar to that of SARS-CoV-2's capsid[7] – constitute the cores of VLPs. Gold is well-known to be suitable for biosensing applications and exhibits unique optical, electronic and catalytic properties.[8–10] Moreover, AuNPs specifically interact with various biomolecules, e.g. antibodies,[11,12] proteins[6,13] and nucleic acids,[14] which constitutes the basis of many virus detection systems. When surface-modified AuNPs are introduced into the solution of protein molecules, coronas rapidly form at their surface through chemical and physical interactions (such as van der Waals forces, hydrogen bonds, coordination, electrostatic or hydrophobic effects, as well as steric hindrance[15]). The coronas of our VLPs are formed by S1 domains of the SARS-CoV-2 spike protein (the "S" protein). This domain was chosen because of its affinity to the ACE2 (Angiotensin Converting Enzyme 2) receptor – which is located at the surface of cells prone to infection by SARS-CoV-2[16] – and mediates the membrane fusion for cell entry.[17,18] The additional advantage of using Au-based VLPs is that they exhibit localized surface plasmon resonance (LSPR), which is a coherent and non-propagating oscillation of free electrons in metallic objects subjected to the interaction with an electromagnetic wave of an appropriate frequency (resonance frequency).[19,20] Usually, it is excited with the use of light of specific wavelength. The oscillation creates a strong electric field around the particle,[21] which can, for example, enhance Raman scattering signals originating from species located in the vicinity of the nanoparticles (leading to the so-called surface-enhanced Raman scattering (SERS)[22]). In the case of protein-covered particles, LSPR – due to its sensitivity to the dielectric environment – can allow detecting specific interactions between proteins and antibodies.[23] Our VLPs are the first to exhibit the LSPR. By utilizing this effect, we were able to visualize the interaction between the VLPs and SARS-CoV-2 monoclonal antibodies (mAbs), which may constitute the basis for the future development of a LSPR-based COVID-19 test. Moreover, functionalization of VLPs with fluorescently-labelled antibodies revealed their potential in studying virus-cell interactions and designing targeted antiviral therapies.

## 2. Results and Discussion

### 2.1. Synthesis and structural characterization of SARS-CoV-2 VLPs

Figure 1 illustrates the general scheme of the synthesis of SARS-CoV-2 VLPs, divided into six steps. Details on the materials, procedures and equipment used in the studies can be found in the Supporting Information (SI) file, section SI1. The synthesized AuNPs were initially coated with cetyltrimethylammonium bromide (CTAB) to prevent their aggregation (Figure 1, step I). Then, they were washed (step II) and CTAB was replaced with bis(p-sulfonatophenyl)phenylphosphine dihydrate dipotassium salt (BSPP) (step III). BSPP is a stabilizing agent[24] and a surfactant necessary for further functionalization of AuNPs with the S1 domain of the S-protein. The optimization of the BSPP coating procedure was performed based on the literature protocols.[25–27] Phosphines bind to Au through a lone phosphorus electron pair.[28] Moreover, the chemical bonds between phosphines and Au are stronger than the electrostatic interactions between citrate or alkyl halides and Au, which allows easy ligand exchange. Additionally, BSPP is less toxic than CTAB, which is an important factor from the perspective of potential biomedical applications. The excess of BSPP was washed out with phosphate-buffered saline (PBS; step IV). Next, the BSPP-coated particles were dispersed in an optimized water solution of S1 domain of the SARS-CoV-2 Spike protein (step V). The free sulphonic groups of BSPP interact with the protein accounting for the interaction between nanoparticles and proteins.[29] Finally, the as-prepared VLPs were incubated in a solution of anti-SARS-CoV-2 Spike S1 mAbs to confirm their biological activity through a specific interaction (step VI).

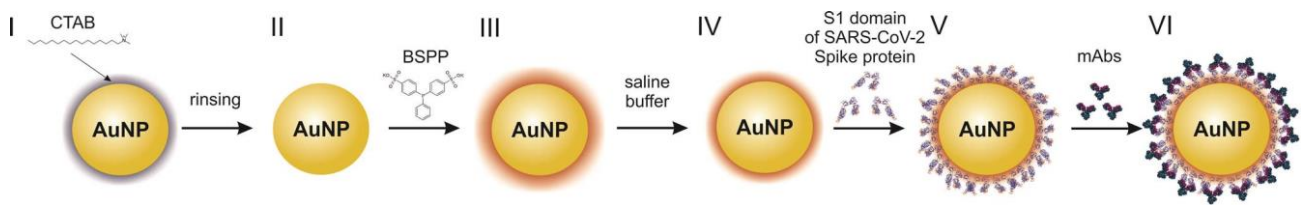

**Figure. 1**. The general scheme of the synthesis of SARS-CoV-2 VLPs consisting of ~100 nm Au cores and S1 protein coronas.

To confirm the presence of S1 domains at the surface of Au NPs, SDS-PAGE electrophoresis was performed. The results are shown in Figure S1. As expected, based on the structural model and the information obtained from the manufacturer, under reducing conditions the protein is characterized by a molecular weight of 92 kDa. For the tested series of VLPs, a band corresponding to the presence of S1 domains appeared at the concentration of about 2 µg/ml, proving its presence at the surface of VLPs. Notably, the intensity of the band was not changing significantly with increasing concentration, which indicates that 2 µg/ml constitutes the amount of protein that is needed to completely cover the surface of AuNPs.

Structural characterization of VLPs was performed using transmission electron microscopy (TEM), scanning electron microscopy (SEM) and dynamic light scattering (DLS) after the most important steps of the VLPs synthesis, i.e. II, IV,V and VI. Additional measurements were carried out after the interaction of VLPs with mAbs (step VI). TEM images are presented in Figures 2(a-d), while the corresponding size distributions obtained from DLS are shown in the insets. SEM results are presented in Figure S2. As can be seen from Figure 2(a), the method used for the synthesis of AuNPs allowed obtaining spherical particles with an average size of approx. 90–100 nm (measured without CTAB) and a narrow size distribution (93.01±1.68 nm; polydispersity index (PDI) of 0.19). The image recorded after exchanging CTAB with BSPP is shown in Figure 2(b), where BSPP is visible as a thin grey shell around the nanoparticle. DLS confirmed the exchange by showing an increase in the average particle size to 103.4±8.03 nm (PDI=0.24). BSPP forms coordination complexes with AuNPs, which renders their high stability and provides protection against aggregation in water solutions.[30] Figure 2(c) shows a TEM micrograph obtained following the attachment of S1 proteins to the AuNPs cores. A "corona" formed by protein molecules, which appear as dark species with a diameter of about 3 nm, is clearly visible. The recorded size-intensity distribution indicates the mean VLP size of about 124.50±18.62 nm (PDI equal to 0.22). The high standard deviation is related to the thickness variation of the corona layer on different particles (ranging from 3 to 10 nm). In order to confirm the biological activity of the synthesized VLPs, they were incubated in a solution of anti-S mAbs. A TEM image obtained after the incubation is presented in Figure 2(d). Compared to pure VLPs, there is a significant difference in the thickness of the corona layer, which increased to 25 nm. The increase was also visible in DLS, which showed an average particle diameter of 175.24±20.41 nm (PDI=0.30). Thus, the performed experiment confirmed biological activity synthesized VLPs.

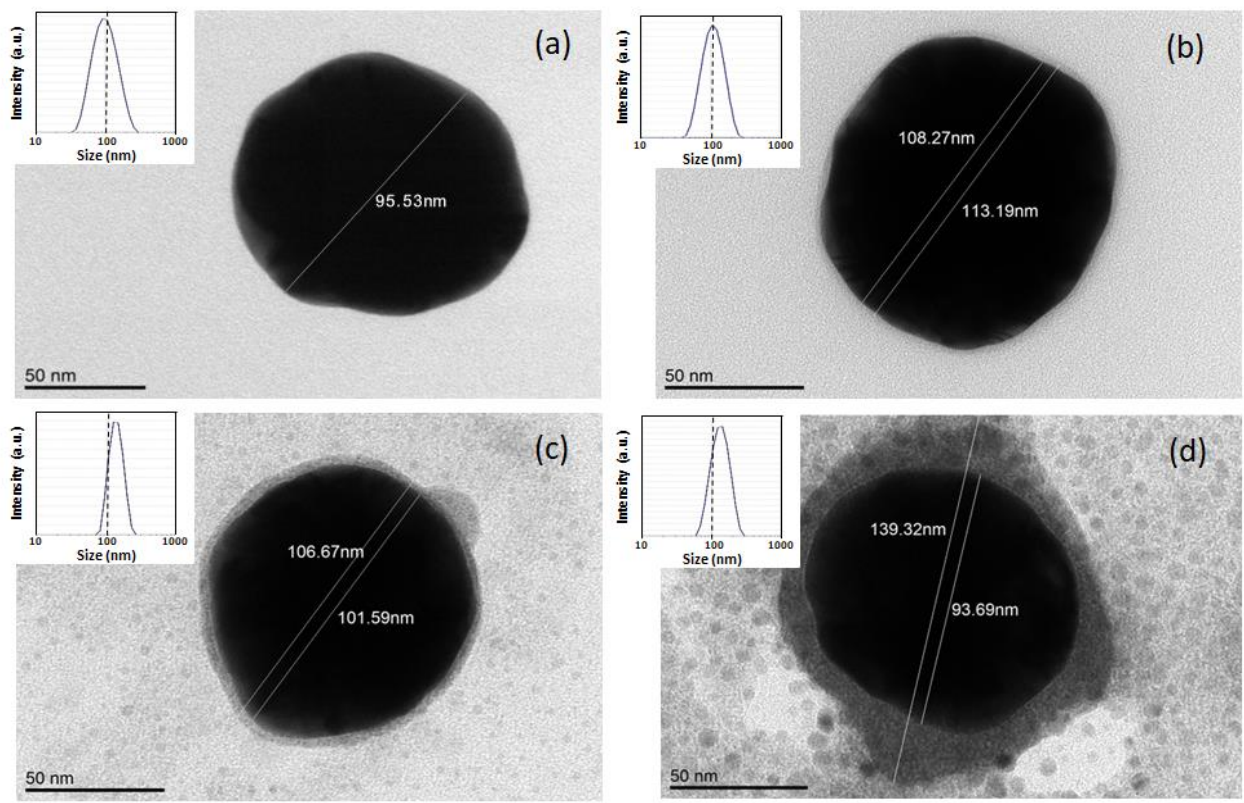

**Figure 2.** TEM images (a-d) and DLS size-intensity distribution histograms (insets) obtained at different stages of SARS-CoV-2 VLPs synthesis: AuNPs after removing the CTAB (a), following the addition of BSPP (b) and after the attachment of S1 proteins (c). (d) shows the VLPs following exposure to SARS-CoV-2 anti-S mAbs.

### 2.2. Surface charge analysis

The change of the surface charge may constitute an indirect proof of successful surface modification. Therefore, the surface charge at different stages of VLPs synthesis was determined from ζ-potential measurements and the results are presented in Figure S3(a). For AuNPs in a water solution of CTAB, a value of 46.85±3.60 mV was determined. The high surface charge of AuNPs/CTAB leads to electrostatic repulsion between particles, which prevents their aggregation. After washing the particles out and dissolving in miliQ water, the surface charge dropped to 23.96±3.80 mV. When the particles were further functionalized with BSPP, sulfonic groups imparted negative charges to the surface of AuNPs,[24] which lead to a similarly high but negative value of the surface charge (equal to -44.42±3.67 mV). The obtained values indicate the importance of prompt replacement of one stabilizing and dispersing agent with the other (in this case CTAB with BSPP) in order to prevent AuNPs aggregation. The formation of a S1-protein corona changed the ζ-potential to -10.00±1.75 mV, which is slightly lower, compared to the average value of native virions of the coronavirus family (-25.68 mV).[31] The VLP/mAb complexes were characterized by a similar surface charge value of -12.25±1.95 mV. These low values obtained for VLPs and VLP/mAb complexes may account for their tendency to aggregate, as observed on SEM images (Figure S2). Most importantly, the evolution of the ζ-potential values at different stages of VLPs synthesis was found to be in line with the morphological changes observed with TEM.

### 2.3. Spectroscopic characterization

Spectroscopic characterization of VLPs was performed using UV-Vis spectroscopy and Raman spectroscopy. The results are presented in Figures 3(a) and 3(b), respectively. The UV-Vis absorption spectrum of pristine AuNPs shows a strong LSPR peak centered at around 567 nm. The position of this peak is related to the size of the particles, their spherical shape and the refractive index of the medium (water). After covering the particles with BSPP, the position of the peak redshifts by 3 nm, which is typical for surface-modified Au species.[32] The resonance shift results from the changes in the dielectric environment of AuNPs related to the attachment of molecules.[33] Following the addition of the S1 proteins, no further shift is observed, which is due to the presence of a relatively thick BSPP layer and the associated negligible influence of additional molecules on the LSPR excited at the Au surface (even a significant change in the dielectric environment taking place several nanometers from the surface does not influence the LSPR in gold[34–37]). Similarly, no shift (within the limit of error, i.e. 1 nm) is observed when combining VLPs with mAbs. However, it has to be noted that the signal recorded for the VLP/mAb complexes is much broader, exhibiting a shoulder at around 800 nm (marked with a red arrow in Figure 3(a)). This change in the signal is related to the formation of VLP/mAb agglomerates of different size that give rise to a family of redshifted adsorption signals (which account for the broadening). Another important aspect, from the point of view of potential applications of VLPs, is their temporal stability. Figure S3(b) reveals that the position of the LSPR peak does not change after 10 days from VLPs preparation and its intensity does not decrease. This confirms that the developed particles are temporarily stable.

Raman spectroscopy measurements allowed us to get insight into the interaction of VLPs with the SARS-CoV-2 anti-S mAbs through the analysis of vibrational signals. The results are shown in Figure 3(b). The presence of LSPR in gold leads to the amplification of Raman signals coming from molecules residing at the surface of AuNPs (the appearance of the SERS effect[38,39]). This facilitates the identification of characteristic vibrational signals of the S1 protein and mAbs, as well as the observation of changes in these signal resulting from the VLP-mAb interaction. For the studies, the solution of VLPs was drop-casted onto a 12-nm-thick Au film deposited on α-Al$_2$O$_3$(0001) (ALO) single-crystal substrate. The selection of the substrate was not incidental, as using a gold film deposited onto a dielectric substrate enhances the SERS effect.[40,41] In addition to VLPs and VLP/mAb complexes, measurements were also performed for clean Au/ALO, as well as AuNPs, AuNPs/BSSP, S1 proteins and mAbs, deposited from solutions onto Au/ALO. In the case of a clean Au/ALO substrate, bands located at 416, 428, 448, 488, 575, 706, 748, 826, 1266, 1370, 1400 and 1657 cm$^{-1}$ were observed. Higher frequency peaks (> 1000 cm$^{-1}$) were expected to appear for Au, while the intensity of peaks positioned at lower frequency (< 1000 cm$^{-1}$) was found to decrease with an increase in thickness of the deposited gold layer (not shown), due to which the peaks were assigned to originate from ALO.[42,43] Only one additional band, located at 496 cm$^{-1}$, was observed after drop-casting pure AuNPs onto this substrate. This peak is most probably related to the presence of sulfur (the S-S vibrational stretching mode[44]) adsorbed on gold from Air. The spectrum obtained for the S1 protein contains numerous bands corresponding to vibrations within different molecular groups.[44–46] The strongest peaks are observed at 494, 525, 542 and 556 cm$^{-1}$ and originate from the stretching of S-S bonds.[44] The bands at 621 and 642 cm$^{-1}$ can be correlated with Phe and C-S stretching modes,[45] respectively, the band at 781 cm$^{-1}$ with the movements of the Tyr residues,[47] while the one at 919 cm$^{-1}$ with the C-C stretching. In the Amide III area, vibrations at 1136 cm$^{-1}$ and 1433 cm$^{-1}$ – coming from CH/CH$_2$ deformations – appear. Further, at 1516 cm$^{-1}$, a signal from Trp is visible. Finally, at 1620 cm$^{-1}$, a peak related to Amide I can be noticed. In the case of SARS-CoV-2 anti-S mAb, which belongs to the immunoglobulins group (IgG), fewer bands were observed: at 491 cm$^{-1}$ (S-S),[48] ~600 cm$^{-1}$ (Phe)[44] and at 1164 cm$^{-1}$ (alkyl C-N vibrations)[47]. The positions of these peaks are similar to those observed for the S1, which is due to the presence of amino acids in the structure of both biomolecules. In addition, vibrations coming from CH/CH$_2$ deformations (1455 cm$^{-1}$)[48], aromatic rings in Trp (at 1544 cm$^{-1}$) and Phe (1600 cm$^{-1}$), as well as the peak related to Amid I band (1630 cm$^{-1}$)[49], were observed. The spectrum of Au/BSPP contains bands at 525, 623 and 697 cm$^{-1}$, coming from the stretching S-S, C-P and C-S modes.[50,51] Due to the fact that BSPP belongs to phosphines, characteristic peaks were also observed at 997, 1032 and 1089 cm$^{-1}$,[52] as well as at 1481 cm$^{-1}$ (CH/CH$_2$ deformations) and 1585 cm$^{-1}$ (benzene rings)[53,54]. The molecule also features two K-O bonds, the peaks of which are visible at 1132 and 1189 cm$^{-1}$, as well as O=S=O groups, giving asymmetric stretching vibrations manifested by a peak located at 1273 cm$^{-1}$.[55] BSPP in combination with AuNPs gives a much clearer signal, due to combined SERS effects from AuNPs and the gold substrate. The spectrum of VLPs contains bands originating from the attached S1 proteins – mainly those located at 493, 530, 621, 630, 780 and 913 cm$^{-1}$. Thanks to the appearance of the SERS effect, two additional bands – at 757 and 977 cm$^{-1}$ – which did not clearly appear in the spectrum of pure S1 proteins, could be observed. The most significant and interesting changes in the spectra occurred after exposing VLPs to mAbs. As the result of the attachment of the antibodies to S1 proteins, three new bands appeared: at ~1068, 1337 and 1557 cm$^{-1}$. The first band lies at the conformationally sensitive region of protein skeletal modes and can be associated with stretching C-N and C-C vibrations in amino acids.[44,48] It is known that protein-antibody interactions involve, in particular, the N-terminal groups of proteins.[56,57] Therefore, the appearance of this band is attributed to the binding mechanism of the antibody to the S1 domain of the Spike protein (its RBD domain)[58]. The second new band, appearing at 1337 cm$^{-1}$, falls within the Amide III region, in which C-H deformation modes are observed. This area is often associated with Trp vibrations, which may also appear in the case of protein/antibody conjugates[44,45,47,48] (as both the stacking interactions from aromatic amino acid rings[59] and Trp are present in the binding of antibodies to proteins[60]). The third new band at 1557 cm$^{-1}$ appears in the Amide II region and comes from the NH$_2$ bending motions,[34,61,62] as well as from Trp[44,45,47,48]. The newly-identified bands related to the specific interaction of VLPs with mAbs may constitute the basis of SERS-based SARS-CoV-2 detection.

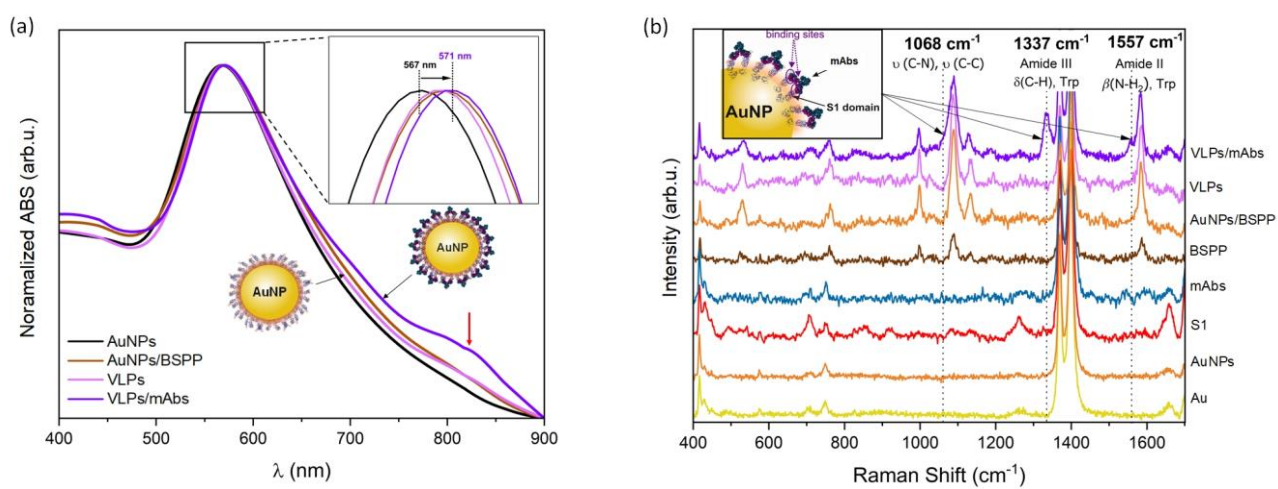

**Figure 3.** UV-Vis (a) and Raman (b) spectra obtained at different stages of SARS-CoV-2 VLPs synthesis and after exposing the VLPs to SAR-CoV-2 anti-S mAbs. UV-Vis reveals the shifts in the LSPR signal. Raman spectroscopy allows observation of peaks related to the interaction of VLPs with mAbs (bands from aromatic amino acids – Trp and Phe).

### 2.4. Fluorescent imaging

Finally, we have evaluated the potential of synthesized VLPs in fluorescent imaging. For this purpose, an Au/ALO substrate with immobilized anti-S mAbs was prepared by irradiating the mAbs solution with UV light for 30 s (total UV intensity on the cuvette: 3 W/cm²) and drop-casting the irradiated solution onto the gold film. UV irradiation is a well-established and effective immobilization procedure, as it promotes the photoreduction of S-S bridges in IgG mAbs through the activation of Trp/Cys-Cys triads, which may either recombine or bind to the neighboring Au species.[48] Next, the synthesized SARS-CoV-2 VLPs were incubated with fluorescent anti-S mAbs (further denoted as FAb350, due to the use of the Alexa-350 fluorescent label) for 5 min and, subsequently, with Au/ALO and mAb/Au/ALO substrates for 30 minutes. After the incubation, each immunocomplex was rinsed 3 times with PBS to detach unbounded FAb350. The confocal microscopy results showed no fluorescence for the Au/ALO and mAb/Au/ALO substrates (Figures 4(a) and 4(b)). For the mAb/Au/ALO incubated with FAb350, which was used as a negative control, because FAb350 should not bind to mAb and give a fluorescent signal, low fluorescence was observed due to a possible non-specific adsorption (Figure 4(c)). However, the mAb/Au/ALO incubated with VLPs/FAb350 gave an intense and uniform fluorescent signal (Figure 4(d)), confirming that the fluorescently-labelled VLPs are still able to react specifically with the antibodies (or other species, such as ACE2 receptors) and, thus, may be used for fluorescent imaging.

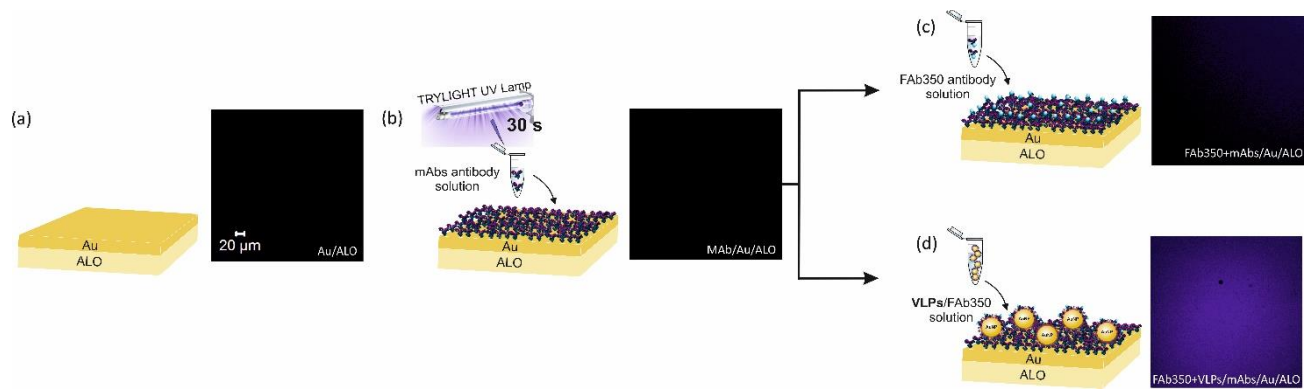

**Figure 4.** Scanning confocal microscopy images obtained for the Au/ALO substrate (a), Au/ALO surface-functionalized with non-fluorescent SARS-CoV-2 Spike S1 mAbs (b), the functionalized surface after incubation with FAb350 (c) and after incubation with VLPs fluorescently-labelled with FAb350 (d).

**Conclusions**

SARS-CoV-2 virus-like particles were synthesized by attaching the S1 domains of viruses' Spike protein to the Au cores constituting the capsids. The particles were characterized with respect to their structure and optical properties, revealing the formation of S1-protein corona and the presence of localized surface plasmon resonance in gold cores. They were also found to preserve their biological activity towards SARS-CoV-2 anti-S monoclonal antibodies. The interaction between the virus-like particles and the antibodies was monitored using Raman spectroscopy and the appearance of three new components, related to the binding of the two species, was observed. The identified characteristic vibrational signals may constitute the basis for Raman-based SARS-CoV-2 infection detection. Moreover, the additionally fluorescently-labelled virus-like particles revealed their potential in fluorescent imaging. Thus, the synthesized particles are suitable for biosensing applications and studying virus-cell interactions. Notably, the developed synthesis procedure, having a relatively universal character, can be potentially applied to fabricate virus-like particles imitating other life-threatening viruses.

**Author contributions**

**W.A.:** Conceptualization, Methodology, Validation, Formal analysis, Investigation, Data Curation, Writing - Original Draft, Visualization; **B.P.:** TEM, SEM – Investigation; **J.L.:** Fluorescence microscopy, electrophoresis – Methodology, Validation, Formal analysis, Investigation, Data Curation, Writing - Original Draft, Visualization; **P.O.:** AuNPs synthesis – Investigation; **J.O.-B.:** AuNPs synthesis – Resources, Supervision; **S.J.:** Supervision; **M.L.:** Conceptualization, Methodology, Resources, Writing - Original Draft, Writing - Review & Editing, Supervision, Project Administration, Funding acquisition.

**Acknowledgments**

The studies were financed through the "Studies on COVID-19" programme from the funds of Adam Mickiewicz University in Poznań, Poland, and – in part – through the LaSensA project carried out under the M-ERA.NET scheme and co-funded by the European Union's Horizon 2020 programme, the Research Council of Lithuania (LMTLT), agreement No. S-M-ERA.NET-21-2, the Saxon State Ministry for Science, Culture and Tourism (Germany) and the National Science Centre of Poland, project No. 2020/02/Y/ST5/00086. The authors would like to thank Zygmunt Miłosz and Szymon Murawka for preparing Au/α-Al$_2$O$_3$(0001) substrates for Raman spectroscopy studies. M.L. would like to additionally express his gratitude to Jan Barciszewski, Krzysztof Lewandowski and Robert Hołyst for critical comments on the results and the manuscript, as well as Katarzyna Lewandowska for proof-checking the language side of the work.

# Supporting Information

# SARS-CoV-2 Virus-Like Particles with Plasmonic Au Cores and S1-Spike Protein Coronas

Weronika Andrzejewska,[1] Barbara Peplińska,[1] Jagoda Litowczenko,[1] Patryk Obstarczyk,[2] Joanna Olesiak-Bańska,[2] Stefan Jurga,[1] Mikołaj Lewandowski[1,*]

[1]*NanoBioMedical Centre, Adam Mickiewicz University, Wszechnicy Piastowskiej 3, 61-614 Poznań, Poland*

[2]*Institute of Advanced Materials, Wroclaw University of Science and Technology, Wybrzeże Wyspiańskiego 2, 50-370 Wrocław, Poland*

[*]*Corresponding author: lewandowski@amu.edu.pl*

*Keywords: SARS-CoV-2, virus-like particles (VLPs), gold, localized surface plasmon resonance (LSPR), Raman spectroscopy, fluorescent imaging*

## SI1. Materials and Methods

### AuNPs

Citrate-stabilized Au nanoparticles (AuNPs; diameter $90 < d \leq 100$ nm) were prepared using the modified methods of Rodríguez-Fernández et al.[63] and Jana et al. (the so-called Turkevich method)[64]. In the first step, gold seeds (d = ~10 nm) were synthesized. In brief, $HAuCl_4$ aqueous solution (25 mL, 0.25 mM) was pre-heated to 100°C and reduced by a fast injection of trisodium citrate aqueous solution (0.5 mL, 34 mM), also pre-heated to ~100°C, under vigorous stirring (1200 rpm). The reaction was carried out in a 50 mL round bottom flask with a reflux condenser which was switched on until the color of the solution changed to ruby red (30 min). The resulting product was purified two times via centrifugation at 18 000 g for 45 min. Then, the supernatant was removed and the pellet was dispersed in the corresponding volume of a Mili-Q grade water. After initial purification, 5 mL of gold seeds were diluted with an equivalent volume of a 0.03 M cetyltrimethylammonium bromide (CTAB) aqueous solution, gently mixed and left overnight under ambient conditions. UV-Vis spectra of as-prepared seed solution revealed localized surface plasmon resonance (LSPR) peak at 523 nm. Subsequently, gold seeds were overgrown to obtain AuNPs with d = ~100 nm. For this purpose, a new growth solution was prepared (0.125 mM HAuCl4, 0.015 M CTAB and 0.5 mM (L)-ascorbic acid) and preheated to 50°C. Afterwards, 0.5 mL of gold seeds were injected into the growth solution, gently mixed by inversion and left undisturbed for 2h. In this process, in addition to AuNPs, also Au species with other shape – such as nanorods or nanospheres – are obtained. Thus, additional purification was carried out following the procedure reported by Jana et al.[64] (in their work, the impurities were separated from the supernatant containing spherical nanoparticles with d = ~100 nm). The resulting solution was purified by centrifugation (3000 g, 5 min), redispersed in water (two times) and, finally, in an 0.01M solution of CTAB (to be stored). All the reagents were used as purchased, without further purification. Water used throughout the experiments was of a Milli-Q grade (Hydrolab DH-0005-UV). Trisodium citrate dihydrate ($Na_3Cit$) ($\geq$ 99 %), gold (III) chloride trihydrate ($HAuCl_4$ x $3H_2O$) ($\geq$ 99.9%), (L)-ascorbic acid (99%) and CTAB ($\geq$ 99%, for biochemistry) were supplied by Sigma Aldrich.

### SARS-CoV-2 Spike S1 proteins and anti-S mAbs

A recombinant SARS-CoV-2 Spike S1 domain C-terminal 6-His tag proteins (Val16-Pro681) and an anti SARS-CoV-2 anti-S monoclonal mouse IgG antibodies (mAbs) were purchased from Bio-Techne, R&D Systems (USA), and dissolved in 0.1 M PBS with the pH of 7.4. The concentration of proteins stock solution was determined by absorbance (A280) using NanoDrop™2000c, with the theoretical molar extinction coefficient calculated using the ProtParam tool (ExPASy).[65] The sodium dodecyl-sulfate polyacrylamide gel electrophoresis (SDS-PAGE) performed for 50 ug/ml of this protein showed the size of about 75 kDa (Figure S1).

**Preparation of SARS-CoV-2 VLPs**

The solution of AuNPs was rinsed by centrifugation and dispersed for stabilization. Then, the CTAB coating was replaced with bis(p-sulfonatophenyl)phenylphosphine dehydrate dipotassium salt (BSPP). For this purpose, AuNPs in 30 mM CTAB solution were heated up to 40ºC in a sonic bath and centrifuged three times at 3000 RCF for 15 min to remove the surfactant. After two centrifugations, the supernatant was removed and AuNPs were dissolved in 1 mL of Milli-Q water. After third centrifugation, AuNPs were dissolved in an earlier prepared BSPP (Merck Milipore) water solution (4 mg, 7.5 mM) and incubated for 6 hours with shaking (850 RPM, 22ºC) in the dark. To remove the excess of BSPP, the sample were centrifuged at 300 RCF for 10 min. The 500 µL of pellet were taken and the rest was re-suspended in additional 500 µL of Milli-Q water. For the preparation of SARS-CoV-2 VLPs, 500 µL of purified recombinant SARS-CoV-2 S1 protein in the concentration of 6.25 µg/mL was added to 500 µL of BSPP-coated AuNPs ($4 \times 10^9$ particles/mL). The mixture was then incubated for 1h at room temperature.

**Absorption spectroscopy (UV-Vis)**

The LSPR signal was recorded for the VLPs solution in standard 1 cm quartz cuvettes (Hellma Analytics), using PerkinElmer Lambda 950 UV-Vis-NIR spectrometer in the range of 300-900 nm (W lamp) with 0.1 nm resolution. The number concentration $N$ of the particles was calculated based on the absorption according to the formula (1):

$$N = \frac{A_{450} \cdot 10^{14}}{d^2 \left(-0.295 + 1.36 \cdot exp^{(-((d-96.8)/78.2))^2}\right)}$$ (1),

where $A_{450}$ is the adsorption measured at 450 nm and $d$ is the diameter of the particles given by the equation (2):

$$d = \frac{\ln(\frac{\lambda_{SPR} - \lambda_0}{K})}{P}$$ (2).

Based on the Mie theory, for nanoparticles with a diameter bigger than 25 nm, the appropriate fit parameters are: $K = 6.53$, P = 0.0216 and $\lambda_0 = 512$ nm.[66]

**Dynamic light scattering (DLS)**

DLS measurements were performed with the use Zetasizer Nano ZS90 (Malvern, UK) equipped with a 633 nm He-Ne laser and a photodiode detector set at the 173º detection angle. The Samples were equilibrated in 25ºC in a standard quartz cuvette. The refractive index of gold was set as 0.2 and the viscosity of the medium to that of water. The Z-average particle diameter with appropriate polydispersity index (PDI) was measured. The average of 10 measurements was used for the analysis.

**ζ-potential measurements**

The ζ-potential measurements were performed using the same instrument as the DLS, with the samples equilibrated in 25ºC in a ζ-potential standard cell.

**Transmission electron microscopy (TEM) and scanning electron microscopy (SEM)**

TEM studies were performed using a JEM-1400 (JEOL Ltd., Japan) microscope with 120 kV operating voltage. 10 µl of samples were dried for 1 min. on the standard 300-mesh Cu grids with carbon Formvar (Agar scientific). After drying, the remaining liquid was removed by touching the grid edge with a low lint paper and stained with 5 µl of 2% uranyl formate solution, which was removed after 1 min. SEM studies were carried out with the use of a JEM 7001F microscope (JEOL Ltd., Japan) with a SEI detector, using 15 kV accelerating voltage.

**Raman Spectroscopy**

Raman spectroscopy measurements were performed using Renishaw's inVia instrument with the 633 nm He-Ne laser. The spectrometer grating conditions were 1800 grooves/mm, acquisition time was 20 s with 1 accumulation. The beam was focused on the samples with a 20× microscope lens with a numerical aperture of 0.4. The measurements were performed in the backscattering geometry with a spectral resolution of $1.0 \text{ cm}^{-1}$. The laser power was adjusted between 50% and 100% ($P_{max}$=17 mW). All the measurements were taken at room temperature, after drying the drop-casted samples onto a $\alpha$-Al$_2$O$_3$(0001) substrate covered by a 12.7 nm-thick Au layer deposited from a crucible with the use of an electron beam (Telemark/PREVAC) under ultra-high vacuum (UHV). In order to extract the Raman signals of interest, the background was subtracted from the acquired raw spectra through the appropriate algorithm, and the data were analyzed using the OriginLab software.

**Fluorescence microscopy**

Fluorescence-activated measurements were performed using a Zeiss LSM 780 confocal laser scanning microscope with a 40× water objective, the excitation laser wavelength of 405 nm and the emission of 441 nm for detecting fluorescence. All the measurements were repeated at least three times to obtain proper statists.

**SDS-PAGE**

Freshly prepared solutions of VLPs prepared using different concentrations of the S1 protein (from 1.6 µg/ml to 6.3 µg/ml) were concentrated by centrifugation at 3000 RCF for 15 min. at room temperature. Then, to extract the proteins from the surface of AuNPs, SDS-PAGE incubation buffer was added. After heating at 95°C, to denature the proteins, and cooling on ice for 2 min, the samples were subjected to a SDS-PAGE gel. Two solutions of S1 protein with the concentrations of 50 µg/ml and 6.3 µg/ml were used as references. SDS-PAGE electrophoresis in reducing conditions was performed by adding incubation buffer to washed VLPs in a 1:1 ratio (15 µl : 15 µL). The samples were subjected to the gel lanes (30 µL/lane) (4–15% Mini-PROTEAN® TGX Stain-Free™ Protein Gels, Bio-Rad). As a reference marker, Precision Plus Protein™ Unstained Standards (Bio-Rad) was used. The experiment was conducted at the constant voltage of 110 V. After electrophoresis, the gel was washed for 5 min. in water, cleaned for 1h in 15% ethanol and in 1% citric acid water solution, washed for 5 min. in water and stained during the night at 4°C in Coomassie® Blue staining reagent (Serva Electrophoresis GmBH, Germany). For visualization, Pharos FX™ Plus Molecular Imager (Bio-Rad) was used.

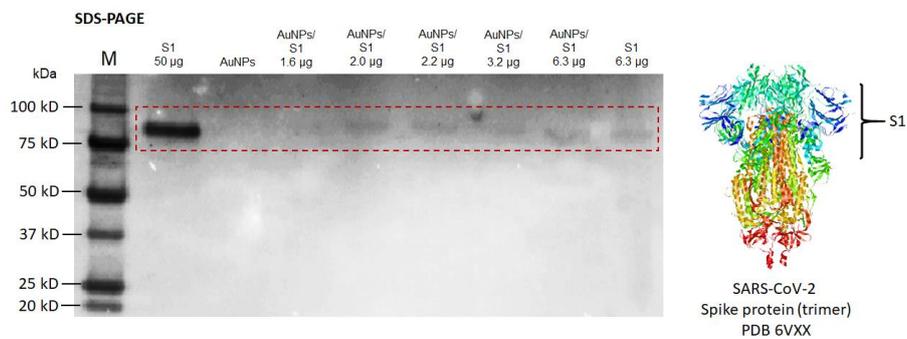

**Figure S1**. (Left) The result of SDS-PAGE electrophoresis obtained for a series of SARS-CoV-2 VLPs synthesized by incubating Au/BSPP cores in solutions of S1 proteins with different concentration per 1 ml (left). (Right) The structure of SARS-CoV-2 Spike protein (PDB 6VXX) [67].

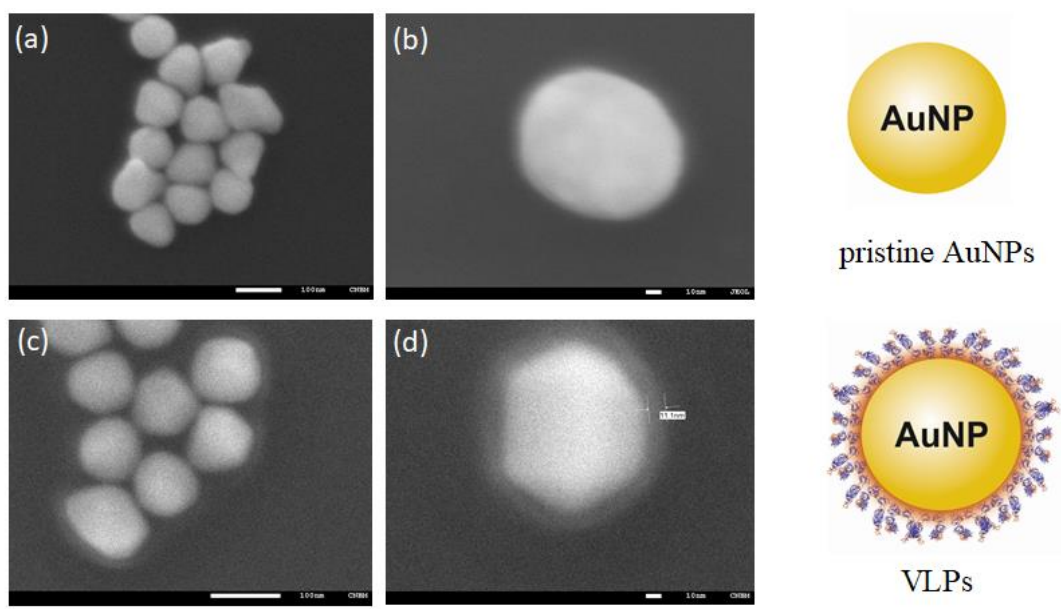

**Figure S2**. SEM images of pristine AuNPs (a,b) and the SARS-CoV-2 VLPs (c,d). The corona of S1 proteins is clearly visible.

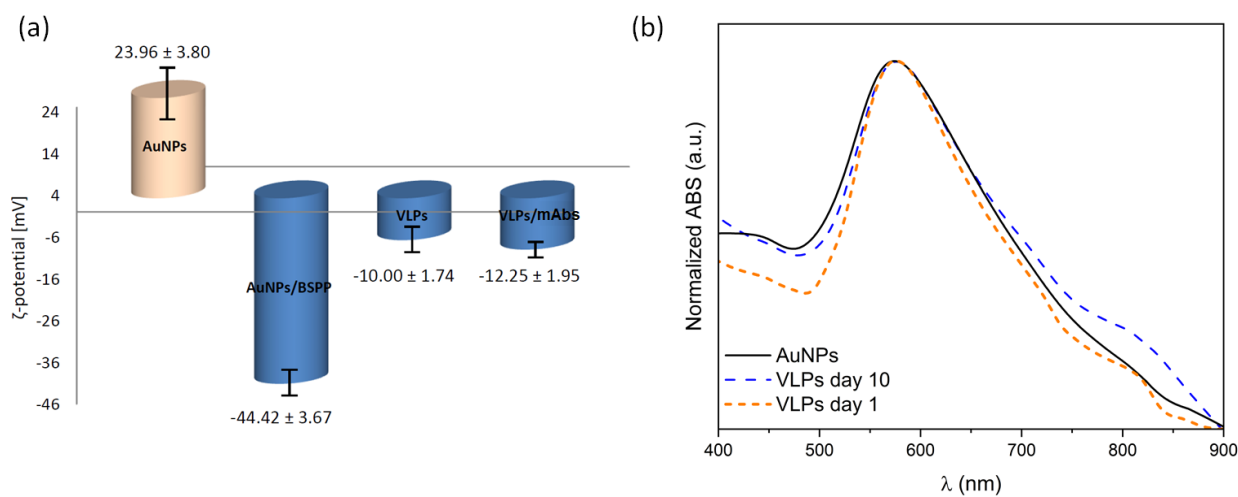

**Figure S3**. (a) ζ-potential values obtained for the studied AuNPs and their conjugates (a). (b) Normalized UV-Vis spectra of pristine AuNPs (black line), freshly-synthesized SARS-CoV-2 VLPs (orange) and the particles after 10 days from the preparation (blue).

**Table S1.** Vibrational modes (Raman spectroscopy peak positions, cm⁻¹) observed for Au/ALO, AuNPs, BSPP, S1, mAb, AuNPs/BSPP, VLPs and VLP/mAb complexes.

| Raman (cm⁻¹) | | | | SERS (cm⁻¹) | | | | Assignment |
|---|---|---|---|---|---|---|---|---|
| Au/ALO | BSPP | S1 | mAbs | AuNPs | AuNPs/BSPP | VLPs | VLP/mAb complexes | |
| 416 | 418 | 416 | 416 | 416 | 418 | 418 | 416 | |
| 428 | | 430 | 430 | 428 | 428 | 428 | 428 | |
| 448 | | 448 | 448 | 448 | 448 | 448 | 448 | |
| 488 | | | | | | | | |
| | | 494 | 491 | 496 | 490 | 493 | 492 | *ν(S-S)str* |
| | 525 | 525 | | | 525 | 530 | 531 | *ν(S-S)str* |
| | | 542 | | | | | | *ν(S-S)str* |
| | | 556 | | | | | | *γ(NH₂), ν(S-S)str* |
| 575 | 577 | 576 | 577 | 575 | 576 | 576 | 576 | |
| | | | 600 | | | | | *Phe* |
| | | 621 | | | | 621 | | *Phe, ν(C-S)str* |
| | 623 | | 630 | | 623 | 630 | 630 | *ν(C-P)* |
| | | 642 | | | | | | *ν(C-S)* |
| | 697 | | 696 | | 692 | | | *ν(C-S)* |
| 706 | | 706 | 710 | 706 | | 708 | 708 | |
| 748 | | 750 | 751 | 748 | | | | |
| | | | | | | 757 | 757 | *Tyr, Trp* |
| | | 781 | | | | 780 | | *Tyr* |
| 826 | 843 | 848 | | 826 | | 832 | 832 | |
| | | 919 | | | | 913 | 913 | *ν(C-C)str* |
| | | | | | | 977 | 977 | *ν(C-C)str* |
| | 997 | | | | 997 | 999 | 996 | *phosphine* |
| | 1032 | 1038 | | | 1032 | 1025 | 1032 | *phosphine, ν(C-* |
| | | | | | | | **1068** | *ν(C-N)str, ν(C-C)str* |
| | 1089 | | | | 1089 | 1089 | | *phosphine* |
| | 1132 | | | | 1132 | 1132 | 1132 | *K-O* |
| | 1189 | | | | 1189 | 1189 | 1189 | *K-O* |
| | | 1136 | | | | | | *Amide III* |
| | | | 1164 | | | | | *alkyl ν(C-N)* |
| 1266 | | 1263 | 1248 | 1266 | | 1266 | 1265 | *Amide III* |
| | 1273 | | | | 1273 | 1273 | | *ν(S=O=S)* |
| | | | | | | | **1337** | **Amide III, δ(CH),** |
| 1370 | | | | 1370 | | | | |
| 1400 | | | | 1400 | | | | |
| | | 1433 | | | | | | *δ(CH,CH₂)* |
| | | 1458 | 1455 | | | | | *δ(CH,CH₂)* |
| | 1481 | | | | 1480 | | | *δ(CH,CH₂)* |
| | | 1516 | 1544 | | | | | *Trp* |
| | | | | | | | **1557** | **Amide II, β(NH₂),** |

| | 1585 | 1594 | 1590 | | 1583 | | | $\nu$(C-C), $\delta$(NH$_2$) |
| | | | 1600 | | | | | Phe |
| | | 1620 | 1630 | | | | 1623 | Amide I |
| 1658 | 1652 | 1659 | 1658 | 1658 | 1655 | 1656 | 1658 | |

The bands were assigned based on the literature (Refs. 38-62) (see the main text for details).